# A Data Model for Integrating Heterogeneous Medical Data in the Health-e-Child Project

Andrew BRANSON, Tamas HAUER, Richard McCLATCHEY, Dmitry ROGULIN and Jetendr SHAMDASANI

*CCS Research Centre, University of the West of England, Bristol, UK*
*{Forename.Surname}@cern.ch*

**Abstract.** There has been much research activity in recent times about providing the data infrastructures needed for the provision of personalised healthcare. In particular the requirement of integrating multiple, potentially distributed, heterogeneous data sources in the medical domain for the use of clinicians has set challenging goals for the healthgrid community. The approach advocated in this paper surrounds the provision of an Integrated Data Model plus links to/from ontologies to homogenize biomedical (from genomic, through cellular, disease, patient and population-related) data in the context of the EC Framework 6 Health-e-Child project. Clinical requirements are identified, the design approach in constructing the model is detailed and the integrated model described in the context of examples taken from that project. Pointers are given to future work relating the model to medical ontologies and challenges to the use of fully integrated models and ontologies are identified.

## 1. Introduction

In recent years, there has been a tremendous increase in the volume and complexity of data available to the medical research community. To enable the use of this knowledge in clinical studies, users generally require an integrated view of medical data across a number of data sources. Clinicians, the end users of medical data analysis systems, are normally unaware of the storage structure and access mechanisms of the underlying data sources. Consequently, they require simplified mechanisms for integrating diverse heterogeneous data sources to derive knowledge about those data in order to have an holistic view of patient information and thereby to deliver personalized healthcare. In this paper we describe a data model and linked ontologies which attempt to provide this holistic view for clinicians which can assist researchers in providing, amongst other things, database query services for clinicians.

The Health-e-Child (HeC) project [1], [2] is an EC Framework Programme 6 Integrated Project that aims to develop a grid-based integrated healthcare platform for paediatrics. It is hoped that using this platform biomedical informaticians will integrate heterogeneous data and perform epidemiological studies across Europe. The resulting Grid-enabled biomedical information platform will be supported by robust search, optimization and matching techniques for information collected in hospitals across

Europe. In particular, paediatricians will be provided with decision support, knowledge discovery and disease modelling applications that will access data in hospitals in the UK, Italy and France, integrated via the Grid.

For economies of scale, reusability, extensibility, and maintainability, HeC is being developed on top of an EGEE/gLite[1] based infrastructure that provides all the common data and computation management services required by the applications. This paper discusses some of the major challenges in bio-medical data integration and indicates how these will be resolved in the HeC system. HeC is presented as an example of how computer science (and, in particular Grid infrastructures) originating from high energy physics can be adapted for use by biomedical informaticians to deliver tangible real-world benefits. The HeC project aims to develop a prototype system which will demonstrate the integration of heterogeneous biomedical data sources over a grid linking multiple hospitals in Italy, the UK and France. In this integration, particular emphasis is put on distinguishing features such as universality of information, person-centricity of information and universality of application leading to the main tenet of the HeC effort: "the integration of information across biomedical abstractions, whereby all layers of biomedical information (i.e. genetic, cell, tissue, organ, individual and population layer) are vertically integrated to provide a unified view of a child's biomedical and clinical condition" [2]. One essential element required for the integration of data across multiple layers of biomedical information is the provision of suitable models for data and information.

The remainder of this paper is organized as follows. Section 2 introduces the requirements behind data integration in HeC and in Section 3 the modelling approach used is considered. Section 4 describes the HeC Integrated Data Model including the approach to metadata specification and then Section 5 outlines the challenges faced in populating that model. The paper closes with conclusions and indications of future work.

## 2. Clinical Requirements

One of the major cornerstones supporting the HeC project goals is the modelling of relevant biomedical data sources. The biomedical information that is managed by HeC spans multiple vertical ranges, comes from different data sources and is possibly distributed and heterogeneous with various levels of semantic content. HeC aims to create a set of models which facilitates the integration of all the available information that supports HeC system components, by providing access to the appropriate information between hospitals and that supports the integration across vertical levels of the medical domain [3].

To be able to combine all sources of data into the integrated view the model of the domain under consideration needs to be established. Such an integrated model must provide clinicians with a coherent view of patients' health and be adaptable to changes in the models of individual sources. Some of the criteria which HeC domain models should satisfy include:
- capturing information specified in clinical protocols;
- supporting high-level applications such as integrated disease modelling, decision support and knowledge discovery;

---

[1] EGEE website - http://www.eu-egee.org/ and gLite Grid middleware website -http://www.glite.org/

- forming the basis of data management in the HeC platform and supporting the clinical queries that are expected to appear in the HeC use-cases;
- being flexible, extendible and able to evolve.

Developing data schema for representing structured medical information has been subject to active research and development during the last decades. HL7 [10] is a standard for information exchange between different medical applications. Although the standard does convey aspects of the hospital process from the financial aspects to the handling of clinical records, HL7 is primarily a messaging standard that enables clinical applications to communicate and exchange medical data. The openEHR Foundation [11] provides a set information models and a terminology based on extendable and reusable formal models of medical concepts (archetypes) and templates allowing the domain models definition independently of terminologies used. Conceptually, the openEHR archetypes are very similar to the metadata definitions in the HeC data model. Although there is no straight correspondence between the openEHR archetypes and the HeC clinical variables types or the openEHR templates and the HeC medical event types but most certainly mappings/interfaces may be provided and the openEHR specifications will be further investigated during the project's lifetime.

## 3. Modelling Approach

To support the HeC objectives, a set of models for representing biomedical information needs to be put in place. As a prerequisite relevant domain knowledge should be reused and at the same time the knowledge base should be aligned to existing data models (such as patient record formats, examination templates, and clinical protocols). Given the heterogeneity and diversity of the large quantity of data that makes up a complete medical record, it is far from easy to capture and align even identical concepts. By collecting common terms that appear in each of the protocols, areas of overlap have been identified. This in turn allowed us to begin to identify key concepts and the relationships between them. Where a form showed data that did not appear in other protocols a decision had to be made as to its inclusion within the model. This is a difficult balancing act since adding everything may produce a structure with a number of redundant sections. On the other hand, there is the danger of missing a piece of data that may prove essential later on. The iterative process of constant review and updating of the models allowed us the flexibility to refine our methods over time. This gradual refinement eventually gave rise to a settled, if not fully integrated, set of components. Finally, the integration of these components within a coherent structure was the last stage in building a conceptual model of HeC domain.

As an initial modelling step, a group of key conceptual entities that inhabit the domain space has been identified including person, hospital, family tree, demographics and several others. The entities chosen represent important roles in the system and also provide an efficient means of sub-dividing complex conceptual relationships into more manageable sections of the overall model. Once such overlapping aspects have been identified and captured in the model, the disease-specific concepts that can add useful extra information should be considered. It is clear that creating new concepts for each and every difference that exists between medical domains would make the conceptual model unduly complicated and perhaps even unworkable. In order to avoid this potential problem it was decided that a set of common medical terms could be produced and that each should serve several roles in the model. In effect, our results amount to identifying reusable patterns of the medical domain which fit the scope of HeC with most also being applicable well beyond.

For instance, much of the clinical data that form the basis of the patients' assessments are acquired by various measurements and represented as physical quantities. Most of these quantities are fully defined in terms of a number (the measurement value) and a suitable measurement unit. Without the (possibly implicit) knowledge of the unit, the quantities cannot be interpreted or compared. Which units are suitable for the attribute of the quantity is determined by its dimension: weight can be measured in kg or pounds but not seconds, etc. The analysis model of the physical quantities is shown on Figure 1.

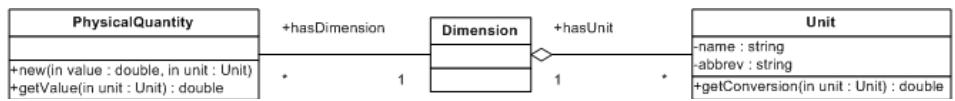

Figure 1. Physical Quantities

A new physical quantity is created using a numerical value and an (existing) Unit instance where the latter automatically establishes the dimension of the quantity (e.g. Joe's height measured [145,<cm>] ). Then the measurement can be queried using other units of the same dimension, in this case one can ask for the value of Joe's height in centimetres or inches. For this to work, units of the same dimension must get converted to each other; this is also represented on the class diagram (see Figure 1).

As physical quantities do not cover all attributes that we need to model in the medical domain similar patterns have been identified for classifications (i.e. corresponding attributes take values from a finite set of discrete possible values, for instance, *Yes/No/Unknown* or *Mild/Moderate/Severe*), clinical observations (for instance, a collection of observations of medical signs with various), free-text annotations etc.

The following section presents how the conceptual models can be harnessed to create a HeC integrated data model and demonstrates how the features captured in the conceptual models are reflected in the data model.

## 4. The Health-e-Child Integrated Data Model

One crucial factor in the creation of integrated heterogeneous systems dealing with changing requirements is the suitability of the underlying technology to allow the evolution of the system [4]. A 'reflective' system utilizes an architecture where implicit system descriptions are instantiated to become explicit so-called "metadata objects" [5]. These implicit system aspects are often fundamental structures and their instantiation as metadata objects serves as the basis for handling changes and extensions to the system, making it somewhat self-describing. Metadata objects are the self-representations of the system describing how its internal elements can be accessed and manipulated. The ability to dynamically augment and re-define system specifications can result in a considerable improvement in flexibility. This leads to dynamically modifiable systems which can adapt and cope with evolving requirements [6]. In this way we can separate the system description in terms of metadata from the particular physical representations of the data and thereby promote ease of integration and querying of the data whilst retaining the ability for the semantics of the system to evolve.

The complexity which arises from the use of diverse distributed data sources in HeC and the anticipated evolution of its medical information led us to the decision to adopt a modelling approach which heavily relies on metadata. In addition the model is enhanced with a semantic layer to facilitate the semantic coherence of the integrated data and to allow linking and reuse of the external medical knowledge. The metadata reveals the structure of the underlying heterogeneous medical data allowing consistent queries across populations of patients and disease types. The semantic layer adds knowledge to this metadata thereby facilitating the resolution of queries that bridge between related concepts. It is this combination of descriptive metadata with system semantics that provides the HeC data model with the ability to be both reactive in terms of the queries generated by user applications and to have the richness to enable integration across heterogeneous data sources. The resulting HeC Integrated Data Model (IDM) constitutes the structures for the representation of data, information and knowledge for the biomedical domain of the HeC (see Figure 2).

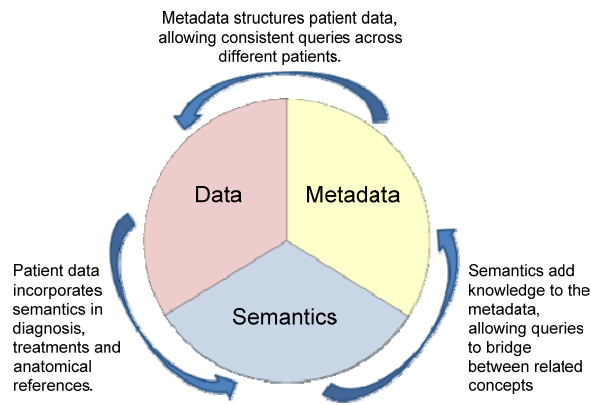

Figure 2. A high-level overview of the HeC IDM

*4.1. HeC IDM:Data*

Conceptually all medical data of the patient can be seen as general patient information (e.g. gender, demographics, family history etc.) with a collection of atomic pieces of data (so-called *clinical variables*) coming from different clinical tests and procedures. The acquisition process is organized as a set of *examinations* that are performed on the patient during visits where each *visit* gives a context/purpose for the examinations. For each patient there can be many visits (e.g. baseline and several follow-ups) at which different examinations (e.g. physical examination, imaging, laboratory test etc.) are performed to acquire different clinical variables (e.g. heart rate, blood pressure, hemoglobin level etc.). Moreover, every visit usually results in setting (or confirming) a diagnosis and/or suggesting some treatment. This information needs to be properly recorded and related to the visit.

Different examinations, diagnosis and treatments are represented as *medical events* i.e. something happened to the patient and was recorded at that particular point in time in the context of some medical interaction. Medical events are always associated with time which can be represented not only as instants (e.g. date of the particular examination) or intervals (e.g. drug prescription) but also relatively to some other event which might be very useful for storing uncertain or incomplete data with respect to the time (e.g. occurrence of some diseases in the past for patient's medical history).

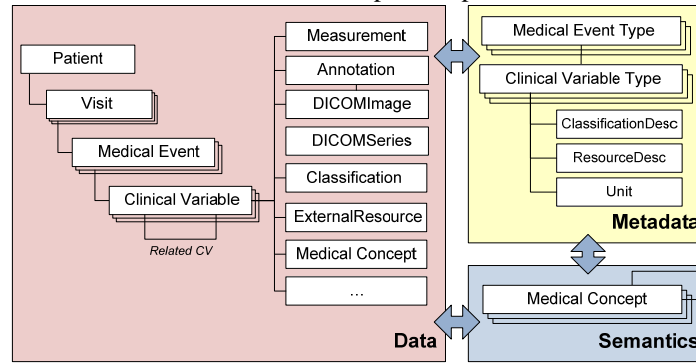

Figure 3. Major entities of HeC IDM

Clinical variables (CV) are grouped within each medical event and represent the actual clinical data as gathered by the HeC protocols. Instead of capturing CVs based on the primitive data types such as 'float', 'integer', 'string' etc. we have identified several major subclasses of CVs based on their clinical meaning and 'essence'. For instance, the aforementioned Physical Quantity pattern can represent any measurement constituted from a numeric value and a unit of measurement. Currently the following categories of CVs have been defined within the HeC domain (see Figure 3):

- **Measurement**: any estimation of the physical quantity (e.g. height, weight, heart rate, right ventricle (RV) volume etc.). It is important to note that each measurement has a numeric value and is associated with a unit of measurement.
- **Annotation**: any free text (e.g. comment, note, explanation etc.). Annotations can be related with any other clinical variables (of different categories) facilitating the efficient storage of any kind of the patient data with the associated annotations.
- **ObservationByClassification**: there are many clinical variables that are assessed based on some classification(s). A classification is a collection of several predefined discrete values which constrains the range of the variable values. The assessment consists of the selection of an element from this collection. For example, the assessment of the severity of RV dilation is based on the selection of one value from the predefined set of strings ("No", "Mild", "Moderate", "Severe").
- **DICOMImage**: a DICOM image can be stored in many different ways (e.g., in the database as BLOB, in the file system, at some URL, on the Grid etc.). The relevant image associated data can be extracted from the image and stored in the database to facilitate the efficient query processing without accessing the image file. DICOMData serves as a container holding the required DICOM metadata. Currently, several attributes are defined mostly for storing the DICOM tags that uniquely identify the DICOM image as well as the study, the series etc. but can be extended according to the emerging requirements.
- **DICOMSeries**: in order to support DICOM temporal series required by the HeC applications (e.g. image registration, segmentation and 3D volume reconstruction tools) DICOMSeries class is introduced. It enables the definition of a series of DICOM images according to the specific purposes and caters for storing DICOM series from the familiar *Patient-Study-Series-Image* hierarchy. Note that DICOM study appears in the IDM as a *medical event*.
- **ExternalResource**: an external resource is defined as any source of the binary data and identified by URI. There is no assumption on the structure of the data in the resource. URI is used to identify the resource. In particular, a file on the Grid is considered as an external resource and the Logical File Name (LFN) that identifies the file on the Grid should be used as URI.
- **MedicalConceptInstance**: any medical event or other clinical variable can be tagged by the medical concept which is defined at the semantic level of the model (see below). For instance, the presence of a particular symptom (e.g. chest pain etc.) is captured through the instantiation of the medical concept representing the symptom as a MedicalConceptInstance object in the model.

*4.2. HeC IDM: Metadata*

There are many noted advantages of the use of meta-data [5], [6] they can improve system interaction and data quality, they can support system and domain integration, and they can enhance system maintenance, analysis and design. In HeC the metadata structures define and describe precisely what data can be stored and how it can be accessed for instance, *medical events* (MET) and *clinical variables* (CVT) *types*, measurement units, classifications and resources descriptions etc. (see Figure 3). *MET*s and *CVT*s are the main organizing entities of the meta-data model. In addition to the kinds of data that can be stored in the model they define the named generic relationships between these kinds and also the grouping of these kinds according to the way the data are collected and managed at the hospitals.

| SEMANTICS | METADATA | DATA |
|---|---|---|
| **Example 1.** *The measurement of Systolic LV volume is to be 30.5 mL/m$^2$* | | |
| | **CVT**: {id="SysLVVol", name="Systolic LV volume", type="Measurement"} <br> **UNIT**: {name="mL/m$^2$"} | **CV**: { type="SysLVVol", value=30.5, unit="mL/m$^2$"} |
| **Example 2.** *Patient X has severe RV dilation* | | |
| | **CVT**: {id="RVDilation", name="RV dilation", type="Classification"} <br> **Classification**: {name="Severity", items="No|Mild|Moderate|Severe"} | **CV**: { type=" RVDilation", value="Severe"} |
| **Example 3.** *Patient Y has a tumour located in Cerebellum* | | |
| Cerebellum $\subseteq$ $\exists$regional_part_of.Brain | **CVT**: {id="TumourLoc", name="Tumour Location", type="MedicalConceptInstance"} | **CV**: { type="TumourLoc", value="fma:Cerebellum"} |

Table 1. Examples of data/metadata/semantics instantiation

Prior to storing any clinical variable in the database its description as CVTs has to be provided. CVTs provide the description of the clinical variables which represent atomic pieces of medical data. Every variable belongs to the particular category (drawing an analogy with UML, ClinicalVariable class is abstract and only its subclasses defined through the categories are instantiable) and the category of the variable is assigned through its corresponding CVT. Every CVT has a human-readable name (e.g. 'Weight', 'RV ejection fraction', 'Haemoglobin measurement' etc.) allowing the schema discovery for users and GUIs (for instance, the names can be used by a query builder tool/GUI to present to the user the kinds of data and to assist in/facilitate the query formulation process). Some examples of the instantiated clinical variable and relevant structures at the metadata and semantics layers are presented in the Table 1.

*4.3. HeC IDM: Semantics*

One of the aims of this model is to represent the clinical data so that it can be populated and interpreted by semantic tools. The key to providing this functionality is the ability to store concepts, then specify typed relationships between these concepts, and between these concepts and the meta-data of the model. 'Concept relates to concept' can then correspond directly to the 'Subject, predicate, object' declarations of RDF[2] and can be recorded at the semantics layer as MedicalConcepts with the URI of the knowledge source. Relating these concepts to CVTs opens up possibilities for browsing and querying software to group together relevant patient data from different patients, visits and medical events. These concepts are grouped together by types which include:
- Anatomical - Body parts, organs and organ components, their relationships and characteristics relating to the three HeC clinical areas.
- Symptoms - Relating to the target diseases, including links to associated diseases.
- Diseases - By family. Differential diagnoses.
- Treatments & Drugs - Drug families, side-effects, etc.

MedicalConcepts are populated by extracting relevant fragments from the existing knowledge repositories and then need to be linked to the CVT as well as the data instances represented in the model as MedicalConceptInstance objects (see Figure 3).

The HeC IDM metadata describes the data structures in the system but does not address how the stored information can be interpreted or how the meaning of (a subset of) the data can be extracted to allow inference of new knowledge, potentially hidden in the data. This interpretation and inference is carried out using the semantic layer which allows information to be integrated or aligned with external data sources or knowledge bases thus permitting knowledge reuse as well as making the knowledge available outside of the project [7]. Consequently, the semantics associated with the data needs to be captured to facilitate the use of integrated information.

**5. Challenges in IDM population**

The clinical data for HeC is specified as a collection of clinical protocols for each candidate disease: specifically a group of paper forms. Each field from each form can be converted directly into a metadata definition, describing the data structure that can be generated from the form, though firstly they should be manipulated to remove presentation and paper artifacts (such as arbitrary length lists becoming variable length arrays, enumerations being derived from long tables with checkboxes etc). Mappings are generated for the metadata to be later used as a guide for data translation.

Using these mappings, data migration becomes simple at the lowest level where one form field becomes one number or string in the metadata, without transformation.

---

[2] Resource Description Framework: http://www.w3.org/RDF/

However, there are many ways of storing the same data, and at this simple level there is no semantic annotation, although this method does suffice for getting data quickly into the database. In our model there are two places where semantic tags can be applied – firstly from the metadata, annotating the description with concepts relating to the nature of the data being measured (e.g. a heart rate measurement linking to the concept of heart rate) and secondly from the data, to instantiate concepts in patient data (e.g. a patient having arthritis in the elbow, requiring a link to the concepts of 'arthritis' and 'elbow joint'). Strings extracted from the form elements can be queried for in ontologies, but this can only ever be a semi-automatic matching process. Although the machine can present best matches, a qualified person must select the correct ones. For querying purposes the semantic part of the database must be sufficiently complete to provide the correct concepts for all of these cases. It would be desirable to link concepts together to a certain degree, but the extent to which this would be useful is, as yet, unknown. The greater the degree of specification in the semantic section, the more powerful the queries that can be written against it, but performance can suffer if it becomes too large. Data migration into a better designed metadata structure is more difficult since text fields from the form must be matched with imported concepts.

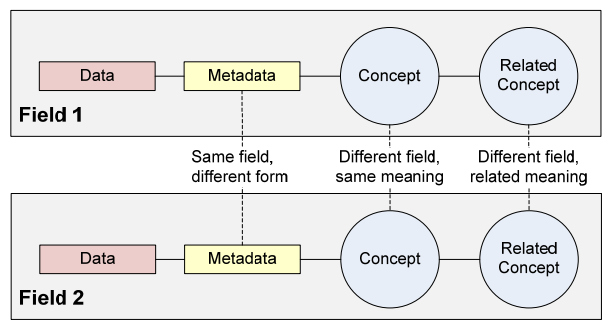

Figure 4. The significance of shared metadata and semantic data

During the population of the description side of the IDM, overlap between semantic and metadata from other areas of the model produces a useful web of links between different fields (see Figure 4). In cases where fields are reused between forms, the overlap will occur in the metadata. For different fields that describe the same concept, they will each link to the same concept. Similarly several metadata elements can be annotated with the same concept (for instance, "heart rate" measurement and "heart murmur" symptom can be related to the "Heart" concept) capturing the fact that, though the fields appear on the different forms, they describe the same concept. Finally, different metadata annotated with different concepts can share a (set of) common related concept(s), facilitating the semantic coherence of the data from different forms.

## 6. Conclusions and Future Work

The approach for data in the HeC project relies on a separation of encoded information into data, metadata and semantics. The data model ensures that all the information recorded can be stored and reused. The metadata model ensured the abstraction required to integrate pieces of data into a coherent whole and to define sufficient description of data elements so that they can be properly interpreted and compared; it is the primary means for data creation and access. To make full use of the information that is captured, one needs sufficient formal semantics associated with the data.

To exploit the full potential of the IDM the metadata and semantics should be populated. The semantic structures add flexibility and descriptive power to the IDM, and, as a consequence, the semantic annotation of clinical protocols becomes crucial. Semantic annotation requires the tagging of data with conceptual knowledge which can be formally represented as, for example, an ontology. The work on semantic annotation using the UMLS Metathesaurus[3] as a primary source of semantics is on-going and the preliminary results suggest good coverage as well as applicability of the existing UMLS annotators in achieving this goal [9].

The current work in the project is evaluating and where necessary extending the use of the IDM in describing the target paediatric diseases in HeC and the use of the model with available medical ontologies to formulate and execute clinician-generated medical queries. The semantic query formulation and enhancement techniques being developed [7] will improve query answering with reasoning capabilities and similarity searches as well as browsing and visualizing patient data [8]. Linkage to external knowledge sources will enrich the data presented about a patient and will enable classification of patients according to various flexible criteria. Finally, application-driven ontology engineering techniques will be the means to investigate how the semantic content can improve the results of high-level applications such as DSS.

The authors would like to acknowledge the support and assistance of all partners in the HeC project, with special thanks to Sonia Zillner and colleagues at Siemens and Tony Solomonides and colleagues at UWE.

---

[3] Unified Medical Language System (UMLS): http://umlsks.nlm.nih.gov